\documentclass{article}

\usepackage{arxiv}

\usepackage[utf8]{inputenc} % allow utf-8 input
\usepackage{graphicx}
\usepackage[T1]{fontenc}    % use 8-bit T1 fonts
\usepackage{hyperref}       % hyperlinks
\usepackage{url}            % simple URL typesetting
\usepackage{booktabs}       % professional-quality tables
\usepackage{amsfonts}       % blackboard math symbols
\usepackage{nicefrac}       % compact symbols for 1/2, etc.
\usepackage{microtype}      % microtypography
\usepackage{lipsum}
\usepackage{amsmath}

\title{MalBERT: Using Transformers for Cybersecurity and Malicious Software Detection}

\author{
  Abir Rahali\\
  \texttt{ear4587@umoncton.ca} \\
\And 
  Moulay A. Akhloufi\\
  \texttt{moulay.akhloufi@umoncton.ca} \\
  
  \And
  Perception, Robotics, and Intelligent Machines Research Group (PRIME)\\
  Dept of Computer Science\\
  Universit\'e de Moncton\\
  Moncton, NB, E1A 3E9, Canada \\
}

\begin{document}
\maketitle

\begin{abstract}
In recent years we have witnessed an increase in cyber threats and malicious software attacks on different platforms with important consequences to persons and businesses. It has become critical to find automated machine learning techniques to proactively defend against malware. Transformers, a category of attention-based deep learning techniques, have recently shown impressive results in solving different tasks mainly related to the field of Natural Language Processing (NLP). In this paper, we propose the use of a Transformers' architecture to automatically detect malicious software. We propose a model based on BERT (Bidirectional Encoder Representations from Transformers) which performs a static analysis on the source code of Android applications using preprocessed features to characterize existing malware and classify it into different representative malware categories. The obtained results are promising and show the high performance obtained by Transformer-based models for malicious software detection.
\end{abstract}

% keywords can be removed

\section{Introduction}
Due to the exponential growth of digitalization usage and easy access to internet technology, cyber threats on information systems such as computers and smartphones increased. Malicious software (malware) is the primary tool used by attackers to perform cyber attacks. Different malware categories such as Trojans, Adwares, and Risktools are expeditiously developed and updated with recent encryption and deformation technologies to become a more severe threat to cyberspace. This rapid development incident often with harmful consequences to different data users at the level of both persons and businesses. For example, IBM reported that the average cost of a data breach is \$3.86 million as of 2020, however, the average time to identify a breach was 207 days \cite{IBM}. As result, IT security professionals and researchers need to update the tools available to automatically detect new malware attacks. 

A series of studies were conducted to solve this issue. Different analysis experiments were used on the static, dynamic and hybrid level \cite{damodaran2017comparison} to extract various types of features such as binaries, permissions, and API calls. The features collected then are passed through detection models developed using machine learning and deep learning tools. The use of deep learning algorithms, for example, helped security specialists to analyze the most complex and targeted attacks. 

The wildly used type is static analysis \cite{pan2020systematic}. It is a known way of identifying malicious applications among benign applications, and this analysis focuses on the source code of software components that may be affected by malware. It is less expensive in terms of resources and time since no need to activate the malware by executing the code to capture the features, it can identify the maliciousness at the code level. For static analysis, there are mainly three practices to detect and classify malware: Permission-based (verify if the requested permissions are needed for the app to ensure the normal behavior on the app access and usage of the user data), Signature-based (identify if the app signature matches one of the malware signatures among the library collected in advance), and specification-based (verify if the app violates the rules set by experts to decide the maliciousness of a program under inspection).

Recent research in deep learning, focus mainly on Transformer-based models such as BERT \cite{devlin2018bert} and XLNet \cite{yang2019xlnet}. These approaches clearly showed impressive results in various state-of-the-art natural language processing (NLP) \cite{tay2020efficient} and computer vision \cite{khan2021transformers} tasks. Thanks to the attention mechanisms layers \cite{vaswani2017attention} added to the encoder-decoder architecture, Transformers can focus on the most important patterns in the data, leading to a remarkable boost in performance.

In this paper, we propose a malware detection approach using a Transformer-based algorithm. We experiment with different Transformer model architectures on our data. The dataset includes 11 different malware categories namely adware, spyware, ransomware, clicker, dropper, downloader, riskware, SMS-sender, horse-trojan, backdoor, and banker \cite{malware_catgories}. Our methodology focuses on the static analysis level on the source code of Android applications, to identify different categories of malware. Indeed, we did not limit the features to permission-based only but considered the whole software code as an important set of feature representation for the analysis. We started with training the model with the features after preprocessing, then a binary classification of the apps to malicious and benign, and finally a cross-category classification at the malware level. 

Our main contributions include:
\begin{enumerate}
\item  We propose a novel approach for malware detection using a  Transformer-based model. According to our best knowledge, this is the first Transformer-based method used for malware classification.

\item  We model Android malware detection as a binary and a multi-label text classification problem and propose a novel feature representation by considering the software applications' source code as a set of features. We apply text preprocessing on these features to keep the important information like permissions, intents, and activities.

\item  We conduct extensive experiments on our preprocessed Android dataset collected from public resources with different category-annotated labels. This preprocessed dataset will be released publicly for the research community.

\end{enumerate}
\section{Background}

Generally, sequence-to-sequence tasks are performed using an encoder-decoder model. And the RNN (Recurrent and Recursive Nets) architectures were the most widely-used architecture for both the encoder and decoder. But these architectures have some limitations.

\subsection{RNN}
RNN or also called sequence modeling like LSTM \cite{hochreiter1997long}, is a family of neural networks for processing sequential data. A recurrent network that maps an input sequence of $x$ values to a corresponding sequence of output $o$ values \cite{hochreiter1997long}. A loss $L$ measures how far each O is from the corresponding training target $y$ \cite{hochreiter1997long}. The loss $L$ internally computes $y = softmax(O)$ and compares this to the target $y$ \cite{hochreiter1997long}. In terms of limitations, RNN based architectures are hard to parallelize because the forward propagation graph is inherently sequential each time step may be computed only after the previous one, where both the runtime and the memory cost are $O(t)$ and cannot be reduced since states computed in the forward pass must be sorted until they are reused during the backward pass.

\subsection{Encoder-Decoder}
The seq2seq model normally has an encoder-decoder architecture \cite{sutskever2014sequence}, composed of an encoder that processes the input sequence and compresses the information into a context vector of a fixed length. This representation is expected to be a good summary of the meaning of the whole source sequence. And a decoder that is initialized with the context vector to emit the transformed output. The mean limitation of the encoder-decoder model is the disadvantage of this fixed-length context vector design is the incapability of remembering long sentences. While the decoder needs different information at different time steps.

\subsection{Transformer}

Thus, in order to solve the limitations of both RNN and endocer-decoder architectures, the authors of Transformers \cite{vaswani2017attention} have proposed a solution. They rely on the seq2seq encoder-decoder by replacing RNN with attention mechanisms. The attention mechanism allows the Transformers to have a very long term memory. A Transformer model can "attend" or "focus" on all the previous tokens that have been generated. The attention mechanism allows the decoder to go back over the entire sentence and selectively extract the information it needs during decoding. Attention gives the decoder access to all the hidden states of the encoder. However, the decoder still has to make a single prediction for the next word, so we can not just pass a whole sequence to it (we have to pass it some kind of synthesis vector). So it asks the decoder to choose which hidden states to use and which to ignore by weighting the hidden states. The decoder then receives a weighted sum of hidden states to use to predict the next word.

In this section, we define the context of the Transformer-based approach by presenting most of the approaches related to the architectures of the approaches. The Transformer in NLP is a new architecture that aims at solving sequence to sequence tasks while easily managing long-range dependencies. The Transformer has been proposed in \cite{vaswani2017attention}. A Transformer is an architecture that avoids recurrence and relies entirely on an attention mechanism to draw global dependencies between input and output. Prior to Transformers, dominant sequence transduction models were based on complex recurrent or convolutional neural networks that include an encoder and a decoder.

Transformers also use an encoder and a decoder, but the elimination of recurrence in favor of attention mechanisms allows for much greater parallelization than methods such as RNNs and CNNs. The transformation is undoubtedly a huge improvement over the seq2seq models based on RNN. But it has its own set of limitations. Attention can only be paid to text strings of fixed length. The text must be divided into a number of segments or pieces before being introduced into the system as input and this causes context fragmentation. For example, BERT \cite{devlin2018bert}, a new linguistic representation model from Google AI, uses pre-training and fine-tuning to create state-of-the-art models for a wide range of tasks. These tasks include question answering systems, sentiment analysis, and linguistic inference. Transfer learning has been used in NLP using pretrained language models that have transformative architectures. 

\begin{figure}[htbp]
	\centering
	\includegraphics[width=8cm]{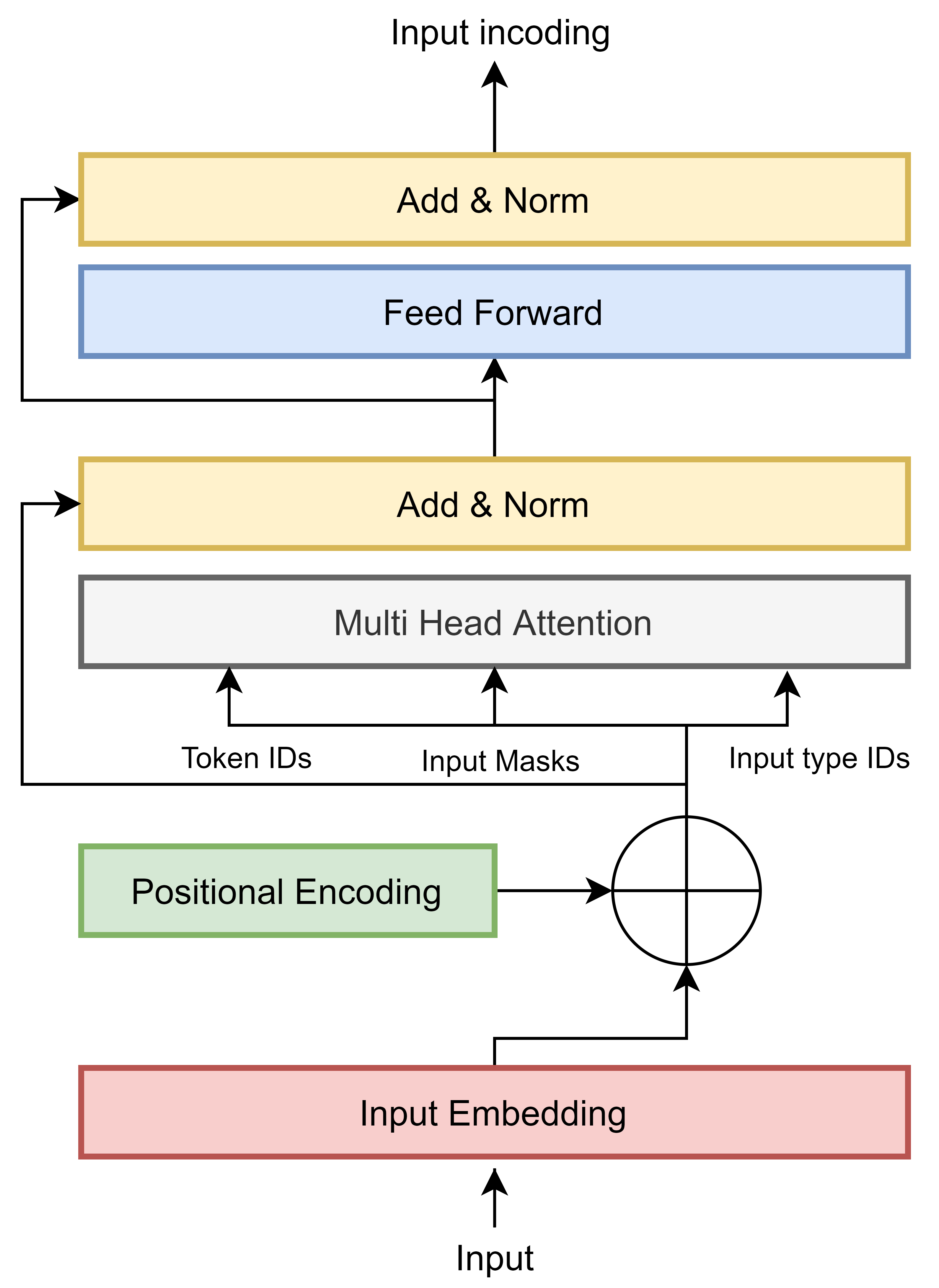}
    \caption{Transformer Encoder Architecture}
    \label{fig0}
\end{figure}

BERT or Bidirectional Encoder Representations from Transformers. Figure \ref{fig0} shows the different layers of the Transformer encoder implemented by BERT. This implementation improves standard Transformers by removing the constraint of unidirectionality through the use of a pre-formed Masked Language Model (MLM) objective  \cite{song2019mass}. The Masked Language Model randomly masks certain elements of the input, and the objective is to predict the original vocabulary identifier of the masked word based solely on its context. Unlike pre-training the left to the right language model, the objective of the MLM allows the representation to merge the left and right context, which allows us to pre-train a deep bidirectional Transformer. In addition to the masked language model, BERT uses a Next Sentence Prediction task that jointly pre-trains the text pair representations. 

Also, RoBERTa \cite{liu2019roberta} an extension of BERT with some modifications in the pre-training procedure of the architecture, uses the same architecture. The modifications include training the model longer, with larger batches, and on a larger amount of data. Also, the predictive objective of the following sentence was deleted. And in order to on longer sequences, the authors applied dynamic modification on the masking scheme \cite{song2019mass} applied to the training data. They also collect a new dataset of comparable size to other privately used datasets to better control the effects of training set size. 

In order to reduce the computation and training time of BERT, \cite{sanh2019distilbert} proposed DistilBERT, a small, fast, cheap, and lightweight Transformer model based on the BERT architecture. This new extension uses knowledge distillation \cite{phuong2019towards} during the pre-training phase in order to reduce the size of a BERT model by 40 \%. To exploit the inductive biases learned by larger models during pre-training, the authors introduce a triple loss combining namely; language modeling loss, distillation loss, and cosine-distance loss.

Another approach based on the Transformer architecture is the autoregressive Transformer. XLNet \cite{yang2019xlnet}, an example of this new implementation, exploits the best of modeling and automatic language encoding while trying to avoid its limitations. Instead of using a fixed forward or backward factorization order as in classical autoregressive models, XLNet maximizes the expected log probability of a sequence with all possible permutations of the factorization order \cite{tay2020efficient}. Also, thanks to the permutation operation usage, the context of each position can be made up of both left and right tokens, which enable XLNet to capture the bidirectional context information of all positions. 

In this paper, we used the presented models in order to evaluate our proposed approach, by conducting a list of experiments on both binary and cross-category malware detection task.

\section{Related works}
Over the past years, various approaches were proposed in deep learning methods to detect malware. While these existing methods are mainly divided between; on one hand the training of recurrent neural networks (RNN), and convolutional neural networks (CNN), on a different set of extracted features, and on another hand the usage of attention mechanisms characteristics for malware classification.

\subsection{Neural network-based approaches}

Among these approaches this study \cite{pei2020amalnet} builds AMalNet, a DN framework to learn multiple integration representations and family assignment with Graph CNN (GCNs) to model high-level graph semantics and use an Independent RNN (IndRNN) to decode deep semantic information. SeqMobile \cite{feng2020seqmobile}, is a behavior-based sequence approach. it uses different recurrent neural networks (RNN). It extracts the semantic feature sequence, which can provide information of certain malicious behaviors, from binary files under a certain time constraint. This paper \cite{niu2020opcode}, presents a new approach based on OpCode-level FCG. The FCG is obtained through static analysis of Operation Code (OpCode) using a Long Short-Term Memory (LSTM). the authors conduct experiments on a dataset on 1,796 Android malware samples classified into two categories and 1,000 benign Android apps. The authors of \cite{jha2020recurrent} focused on step size as an important factor in relation to input size using RNN. They tested the model with three different feature vectors (hot-coding feature vector, random feature vector and Word2Vec feature vector) using hyper parameters. \cite{nait2020intelligent} transform the android package kit (APK) file into a lightweight RGB image using a predefined dictionary and intelligent mapping, then apply a CNN on the obtained images for malware family classification. Multiple other examples of DNN based approach \cite{wang2020review} and \cite{mercaldo2020deep}, have been developed, with varying the feature extraction, selection, and representation methods in the aim of boosting the detection results.

\subsection{Approaches using Attention Mechanisms}

While our approach is, to the best of found knowledge, the first study to implement Transformers directly on software application and preprocessing features like text to detect malware, few approaches have attempted first steps towards using new methods such as attention mechanisms.

For example, \cite{chen2020slam} propose SLAM a malware detection framework build based on the characteristics of the attention mechanism and the sliding window method. It use a feature extraction method of the API execution sequence according to its semantics. \cite{ganesan2020robust} uses of an residual attention based mechanism, based on this study the proposed method outperformed traditional CNN models. Similarly, \cite{zhang2020ransomware} propose a static analysis framework based on a set of  N-gram opcodes sequence patches with a self-attention based CNN named SA-CNN. Another example proposed by \cite{yakura2018malware}, applies a CNN with an attention mechanism to images converted from binary datasets, by calculating an attention map to extract characteristic byte sequence. The distinction of regions in the attention map shows regions having higher importance for classification in the image. 
 
While in our work, we propose a noval approach, we used BERT to better detect malware, we refined the pre-trained model to efficiently learn representations of source code language syntax and semantics. Our Context-Aware network learns contextual characteristics from a natural language sentences perspective thanks to the attention mechanism layers in the Transformer-based architecture.

\section{Methodology}
\begin{figure}[htbp]
	\centering
	\includegraphics[width=8cm]{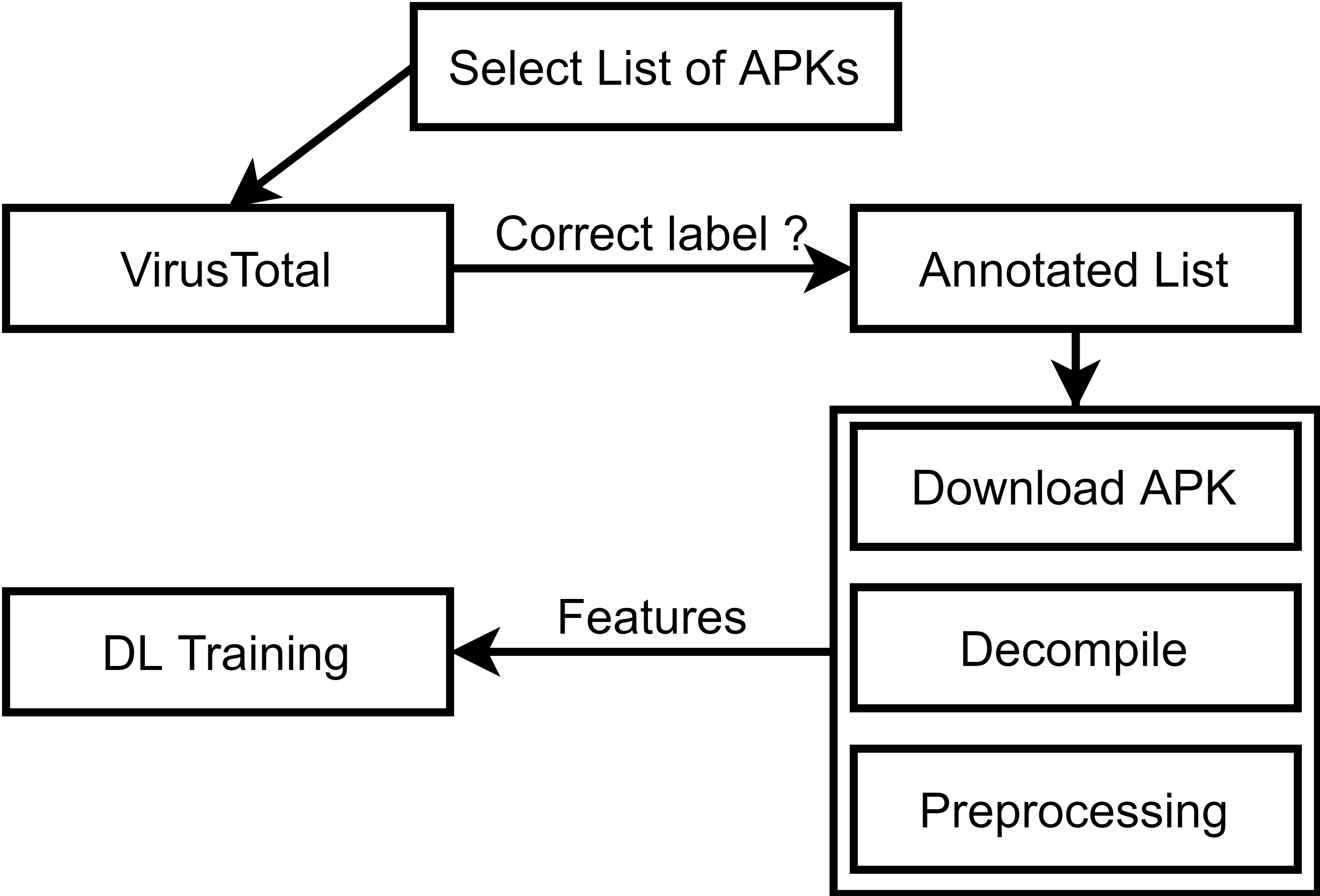}
    \caption{Overview of the Methodology Steps}
    \label{fig1}
\end{figure}

This section explains the overall process of malware detection. The core idea of this work is to create, a malware detection framework using a Transformer-based approach. To reach this goal, we conducted a static analysis on the collected corpora from a natural language sentences perspective. So, we need a dataset including source code files and different categories of malware types. Figure \ref{fig1} explains the logical flow of our Android malware detection. This  Process is mainly divided into 4 phases. First, the Android files collection, then the Decompilation phase of the APK files, Feature Mining, and finally Deep Learning (DL) models training experiments.

\subsection{Data Collection}
We collected the Android applications from the Androzoo public dataset. Androzoo, one of the stae of the art android malware dataset \cite{Allix:2016:ACM:2901739.2903508}, is a growing collection of Android applications from several sources, including the official Google Play app market. It currently contains 13,320,014 different APKs, each of which has been analyzed by dozens of different antivirus products to find out which applications are detected as malware. This public data is up to date with weekly analysis on the samples \cite{liu2020androzooopen}. The data is labeled based on these analyses into malware and benign, and different malware categories and families. 
\begin{figure}[h]
	\centering
	\includegraphics[width=9cm]{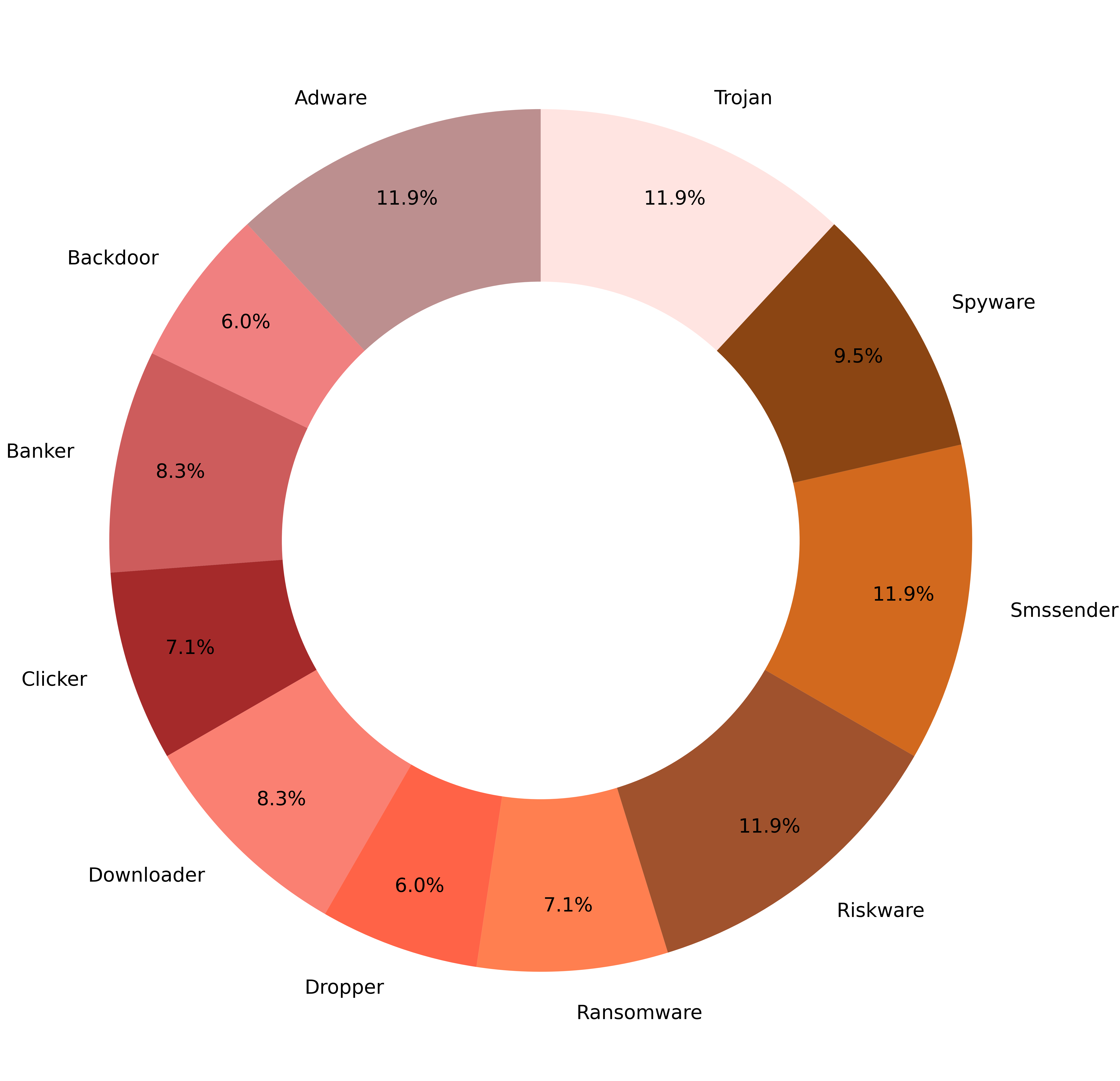}
    \caption{Android Malware Categories}
    \label{fig10}
\end{figure}

Based on state-of-the-art taxonomies for Android malware categories \cite{ismail2017general} \& \cite{chouhan2017preface} We selected 11 categories \ref{fig10} namely; adware (displays advertising and entice a user to install it on their device), spyware (installs itself on the user device with the aim of collecting and transferring information without the user is aware of it), ransomware (takes personal data hostage), clicker (a type of trojan that performs a form of ad fraud. These “clickers” continuously make connections to websites, consequently awarding threat actors with revenue on pay-per-click bases), dropper (a syringe program or dropper virus, is a computer program created to install malicious software on a target system), downloader (a type of Trojan horse that downloads and installs malicious files), riskware (a software whose installation can represent a risk for the security of the computer, but however, not inevitably), SMS-sender (presents itself as a regular SMS messaging application and uses its basic permissions to send/receive short messages), horse-trojan (is designed to damage, disrupt, steal, or in general inflict some other harmful action on your data or network), backdoor (when introduced into the device, usually without the user's knowledge, turns the software into a Trojan horse, and banker (is designed to steal data from users' online bank accounts as well as data from online payment systems and plastic card systems). We select the list of APKs to download based on the recent creation and analysis date, then re-analyze this list with VirusTotal \cite{total2012virustotal}, to finally create our dataset list including 12,000 benign apps and 10,000 malware apps. 

\subsection{Preprocessing and Feature representations}
Once the list of APKs is defined, we write a script to download the files. Then, we decompiled the downloaded APKs using Jdax \cite{jadx}, which creates folders of the apps' files \cite{harrand2020java}. We extracted the AndroidManifest.xml file from each sample. The manifest presents essential information about the application to the Android system, information the system must have before it can run any of the application's code, including the list of permissions, the activities, services, broadcast receivers, content providers, the version, and the meta-data. These files are then treated as text files and passed through the preprocessing phase, in this step and to conserve the important information about the features, we apply specific cleaning of the not important, mostly repeated words, in the code. We manually analyzed different examples and created a list of words and expressions that do not provide additional info, so the cleaning included lexicon removal, punctuation removal and we conserved the digits and the cases of the characters. The purpose of the preprocessing is to reduce the size of the input. The final dataset format has 4 columns, the ID column, represented by the APK hash name, the Text column representing the Manifest files after preprocessing, the Label column, a binary format equal to 1 if the app is malware and 0 if not, and finally the Category column representing the malware type name (exp: adware).

\subsection{Proposed Approach}

\begin{figure}[htbp]
	\centering
	\includegraphics[width=8cm]{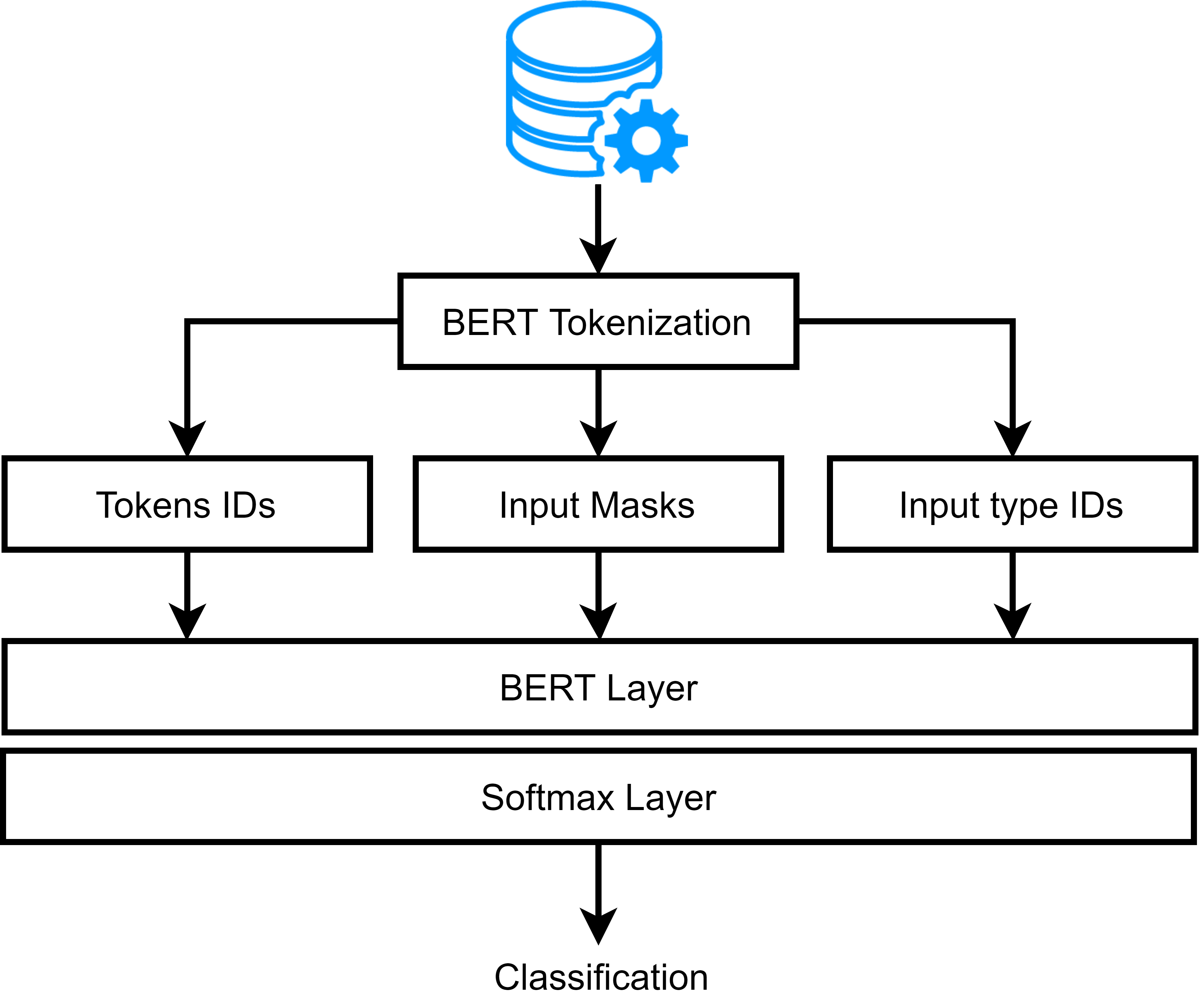}
    \caption{BERT model fine-tuning process}
    \label{fig2}
\end{figure}

Once the data is created and annotated in the right format,
we split the data into train and test. We conduct all our experiments using BERT, we fine-tuned it on our train dataset. We fixed the hyperparameters based on each classification type. We train BERT to predict Malware/Benign (i.e., binary classification) for each sample, then, to predict the categories of malware (i.e, multi-classification). The Transformer architecture has specific input formatting steps including the creation of special tokens and ids.  We use the Transformers implementation of the hugging face library \cite{wolf2019huggingface} for the binary classification of Android applications. Only the Transformer architecture, layers, and weights are implemented, while all data formatting must be done beforehand to be compatible with the Transformer. While most pre-trained Transformers have essentially the same steps. Here we test this approach with BERT. Figure \ref{fig2} gives a detailed overview of our approach.

%\clearpage
\section{Experimental results}
To test the proposed approach, we evaluate it in terms
of three main aspects: (1) the profitability on large and recent categorical datasets, (2) the feature representation ability for information context extraction from android apps, and (3) the performance compared to the state-of-the-art approaches.
\subsection{Experiments setup}
In our model implementation, we used Adam optimizer and set its hyperparameters as follow; the  learning rate $\alpha$ to 2e-5 for binary classification and 1e-3 for multi-classification, $\alpha$1=0,9 and $\alpha$2=0,999. We set the max sequence size to 512 and the training batch size to 32.

\subsection{Baselines}
We compare the proposed model with several state-of-the-art methods including LSTM \cite{staudemeyer2019understanding}. We selected this sequence model based on its performance in the reviewed studies. And, XLNet \cite{yang2019xlnet}, RoBERTa \cite{liu2019roberta}, and DistilBERT \cite{sanh2019distilbert} as Transformer based baselines to compare their performance to BERT.

\subsection{Evaluation Metrics}
To evaluate the effectiveness of the model and avoid the contingency caused by the partitioning of the train and test sets, we select three metrics to evaluate the model (Accuracy, MCC, and Loss), which are commonly used in classification problems. We define the following numbers; TP (true positives), FN (false negatives), FP (false positives), and TN (true negatives).
\begin{enumerate}
    \item ACC: The accuracy is a metric for evaluating classification models. It is equal to the number of correct predictions divided by the total number of predictions, as equation 1 present:
    \begin{equation}
        Accuracy = \frac{TP+TF}{TP+TF+FP+FN}
    \end{equation}
    
    \item MCC: The Matthews Correlation Coefficient (MCC) is bast used for binary classification with an unbalanced dataset. It has a range of -1 to +1. We chose MCC over F1-score for binary classification as recommended in this study \cite{chicco2020advantages}. MCC equation is defined as fellow :
    
    \begin{equation}
    \fontsize{6}{8}
        {\displaystyle {\text{MCC}}={\frac {{\mathit {TP}}\times {\mathit {TN}}-{\mathit {FP}}\times {\mathit {FN}}}{\sqrt {({\mathit {TP}}+{\mathit {FP}})({\mathit {TP}}+{\mathit {FN}})({\mathit {TN}}+{\mathit {FP}})({\mathit {TN}}+{\mathit {FN}})}}}}
    \end{equation}
    
    \item F1: The formula for the standard F1-score is the harmonic mean of the precision and recall. A perfect model has an F-score of 1. We used macro averaging in the results we presented in this paper. The formula of F1-score is defined as follow:
    \begin{equation}
        F1 =  \frac{2*TP}{2*TP+FP+FN}
    \end{equation}
    
    \item Loss: We used the cross-entropy loss, or log loss \cite{zhang2018generalized}. It measures the performance of a classification model whose output is a probability value between 0 and 1. Cross-entropy loss increases as the predicted probability diverge from the actual label. The cross-entropy $\displaystyle H(p,q)$ of the probability distribution ${\displaystyle p}$ relative to a probability distribution ${\displaystyle q}$ is given as:
    \begin{equation}
        {\displaystyle H(p,q)=-\sum _{x\in {\mathcal {X}}}p(x)\,\log q(x)}	
    \end{equation}

\end{enumerate}

\subsection{Results and Analysis}

We conducted the experiments on the preprocessed dataset. Fine-tuning the pre-trained models, clearly gave the highest accuracy results for this classification task compared to the LSTM baseline. The best classification model is BERT. The test metrics results of Table \ref{tab1} show that each Transformer learns differently depending on each architecture. The results in Table \ref{tab1} and Table \ref{tab2} prove that BERT outperformed the other baseline models in both binary and multi-classification malware detection. For BERT, the best learning rate shows that only two epochs are required before the loss starts to increase. Our fine-tuning with the training set included changing the hyperparameters to boost the results. To evaluate the final results, we used different evaluation metrics. The pretrained models achieved good results overall, but BERT obtained the best performance in both tasks.

\begin{table}[htbp]
\centering
\begin{tabular}{cclcc}
\hline
\textit{\textbf{Algos}} & \textit{\textbf{ACC}} & \textit{\textbf{F1}} & \textit{\textbf{Loss}} & \textit{\textbf{MCC}} \\ \hline
LSTM                    & 0.9405                & 0.9382               & 0.1521                 & 0.9077                \\
XLNet                   & 0.9579                & 0.9549               & 0.1461                 & 0.9164                \\
RoBERTA                 & 0.9533                & \textbf{0.9499}               & 0.1430                 & 0.9070                \\
DistilBERT              & 0.9542                & 0.9542               & 0.1305                 & 0.9087                \\ \hline
BERT                    & \textbf{0.9761}                & 0.9547               & \textbf{0.1274}                 & \textbf{0.9559}                \\ \hline
\end{tabular}
\caption{Detection results using the feature representation approach across difference networks on the test dataset for both binary classification.}
\label{tab1}
\end{table}

\begin{table}[htbp]
\centering
\begin{tabular}{cccc}
\hline
\textit{\textbf{Algos}} & \textbf{ACC} & \textit{\textbf{F1}} & \textbf{Loss} \\ \hline
LSTM                    & 0.8507       & \textbf{0.7816}               & 0.3521        \\
XLNet                   & 0.6232       & 0.7448               & 0.8106        \\
RoBERTA                 & 0.6491       & 0.7670               & 0.7611        \\
DistilBERT              & 0.5981       & 0.7274               & 0.8505        \\ \hline
BERT                    & \textbf{0.9102}       & 0.7804               & \textbf{0.2214}        \\ \hline
\end{tabular}
\caption{Detection results using the feature representation approach across difference networks on the test dataset for cross-category malware classification.}
\label{tab2}
\end{table}

%\clearpage
\section{Conclusion and future works}
This paper studies the challenge of malware classification using our novel approach for Transformer-based malware detection.  We detailed the malware text classification methodology and used it for feature representation. The BERT based model achieved high accuracy results for both binary and cross-category classification, compared to the other baseline pre-trained language models. The results from the experiments show that the best binary accuracy is 0.9761 and for the multi-classification, it is 0.9102. We can conclude that the proposed approach's results of feature representation as text input for a Transformer-based model, are very good. So, the implementation of pre-trained linguistic models based on Transformer architectures in cybersecurity tasks can outperform standard RNN models like LSTM, when applied on a state-of-the-art dataset like Androzoo. Future work will therefore consist of testing other models, testing other types of data, and creating an API to detect malware in new applications.

\bibliographystyle{unsrt}  
\bibliography{references}  %%% Remove comment to use the external .bib file (using bibtex).

\end{document}